\documentclass{jfm}
\usepackage{natbib}
\usepackage{epsfig}
\usepackage{amssymb}
\usepackage{color}

\def\begineq{\begin{equation}}
\def\endeq{\end{equation}}
\def\be{\begin{equation}}
\def\ee{\end{equation}}

\definecolor{gray}{rgb}{0.6,0.6,0.6}
\title[Asymmetry of temporal cross-correlations in turbulent shear flows]
{Asymmetry of temporal cross-correlations in turbulent shear flows}
\author[A. Jachens, J. Schumacher, B. Eckhardt, K. Knobloch and H. H. Fernholz]%
{A.\ns J\ls A\ls C\ls H\ls E\ls N\ls S,$^1$
   \ns J.\ns S\ls C\ls H\ls U\ls M\ls A\ls C\ls H\ls E\ls R,$^1$ 
   \ns B.\ns E\ls C\ls K\ls H\ls A\ls R\ls D\ls T,$^1$
   \ns K.\ns K\ls N\ls O\ls B\ls L\ls O\ls C\ls H$\,^2$
   \ns \and \ns H.\ns H.\ns F\ls E\ls R\ls N\ls H\ls O\ls L\ls Z$\,^2$}
\affiliation{$^1\,$Fachbereich Physik, Philipps-Universit\"at, D-35032 Marburg, Germany\\
$^2\,$Hermann-F\"ottinger-Institut, Technische Universit\"at, D-10623 Berlin, Germany}
\pubyear{2004}
\volume{999}
\pagerange{1--10}
\date{06 July 2005 and in revised form ??}
\begin{document}
\maketitle
\begin{abstract}
We investigate spatial and temporal cross-correlations between
streamwise and normal velocity components  
in three shear flows: a low-dimensional model for vortex-streak 
interactions, direct numerical simulations for a nearly homogeneous 
shear flow and experimental data for a turbulent boundary layer.
A driving of streamwise streaks by streamwise vortices gives rise to a temporal asymmetry
in the short time correlation. Close to the wall or the 
bounding surface in the free-slip situations, this asymmetry is 
identified. Further away from the boundaries the 
asymmetry becomes weaker and changes character, indicating
the prevalence of other processes.
The systematic variation of the asymmetry measure may be used 
as a complementary indicator to separate 
different layers in turbulent shear flows. 
The location of the extrema at different streamwise displacements 
can be used to read off the mean advection speed; it differs
from the mean streamwise velocity because of asymmetries in
the normal extension of the structures.
\end{abstract}

\section{Introduction}
Coherent structures are very effective
in transporting momentum across velocity gradients and thus
contribute significantly to frictional drag in turbulent flows.
Depending on the type of flow and the position of the layer being
studied, different kinds of structures can be identified 
(Robinson, 1991; Panton 2001). In wall bounded shear flows,
Robinson (1991) describes a dominance of streamwise vortices close
to the walls and horseshoe-like structures in the outer region. For the
intermediate region one might imagine a gradual transition in 
relative weight from one to the other. In transitional internal flows at
low to intermediate Reynolds numbers streamwise vortices 
and streaks are also present
(Eggels {\it et al.} 1994; Hof {\it et al.} 2004; Grossmann 2000),
and a complete self-regenerating
cycle for the dynamics, in which 
vortices drive streaks which then generate vortices 
through a shear instability,
has been proposed (Waleffe 1997). 
The relation between internal and wall-bounded flow
situations has been established through simulations
in laterally confined geometries which show a similar dynamical
behaviour (Hamilton {\it et al.} 1995).
The presence of vortices and streaks in flows with homogeneous shear
(Kida \& Tanaka 1994; Schumacher \& Eckhardt 2001) and calculations
within rapid distortion theory (Nazarenko et al 2000) highlight the 
significance of the background shear for their evolution and dynamics.

Streamwise streaks result from the mixing of fluid across the shear
gradient as induced, for instance, by streamwise vortices. This
is a linear process that suggests a causal relation between their
appearance: the vortices have to be there first, and can then 
be followed by the streaks. 
With the pointwise measurements in boundary
layers in mind, we take velocity components as indicators for the
structures: the streamwise 
turbulent velocity component $u$ for the streaks and
 the wall-normal or shear component $v$ for the vortices. 
The temporal correlation can then be verified 
for linear models of the vortex-streak interaction by direct 
calculation (Eckhardt \& Pandit 2003). In particular,
the model shows that the temporal cross-correlation
function $C_{vu}(\Delta t)=\langle u(t+\Delta t)v(t)\rangle_t$ 
will be asymmetric, and it will have its extremum at a finite
time-delay $\Delta t$. The vortex as measured by $v$ has a chance 
to influence the streak in $u$ for $\Delta t>0$,
but not for $\Delta t<0$. 
The question we address here is the 
extent to which this causal relation is reflected in an 
appropriate temporal correlation function in fully developed
turbulent flows. 

The correlations we are interested in can be obtained from
two-point data, from measurements displaced in space or time or both. 
Apparently, the first such data were obtained by 
Blackwelder \& Kovasznay (1972) in a turbulent boundary layer. 
Their correlations for $y/\delta\approx 0.2$
\footnote{Throughout this work, the boundary layer 
thickness $\delta$ is defined 
as the distance between the wall and the height $y$ where 
$U=0.99 U_{\infty}$ with the free-stream 
velocity $U_{\infty}$.} 
show a weak asymmetry of under 
$\Delta t\to -\Delta t$. 
Later, Blackwelder \& Eckelmann (1978),
studied cross-correlations between normal derivatives of the
velocity components, and confirmed the asymmetry, as well as a shift in
the maximum towards a positive time shift. 
Lagrangian studies of this cross-correlation along particle paths
also show this asymmetry (Pope 2002).

Fully turbulent dynamics differs from the linear model with stochastic
forcing by the presence of nonlinearities and a
self-consistent generation of the turbulent fluctuations. In addition,
the self-sustaining cycle proposed by Waleffe (1997) 
for the complete dynamics
consists not only of the non-normal amplification but also of
a linear instability that generates normal vortices which are then
tilted into the streamwise direction by the main flow. The asymmetry
is a property of the first part of the cycle, but not
of the second. The correlation function of the full cycle will
thus be a superposition of the contributions from the both parts,
and the asymmetry will show up if the statistical weight of
the second part is smaller than that of the first part.

With this in mind, we want to turn to the analysis of the 
cross-correlation functions of the turbulent velocity 
components for evidence of the dynamical processes underlying the
self-sustaining mechanism for the formation of coherent structures. 
We will study in detail data from a low-dimensional model, 
from direct numerical simulations (DNS)
of shear flows and from measurements with 
hot-wire probes in a boundary layer.
The model allows for a detailed tracking of the dynamics of the various
contributions to the spatial and temporal correlations. 
The DNS allows for an extension to two-point cross-correlations 
in space and time since they do not have to rely on 
Taylor's frozen flow hypothesis. Finally, the experimental data, although 
restricted to time asymmetry, allow for much higher Reynolds numbers
and for a systematic study of the dependence of the 
correlation functions on the distance from the
wall. 

We take coordinates with $x$ in streamwise, $y$ in wall-normal and $z$
in spanwise directions. The quantity we focus on is the correlation function
between the normal velocity component $v$ and the streamwise component $u$,
displaced in the streamwise direction by $\Delta x$ and in time by $\Delta t$,
\begin{equation}
C_{vu}(\Delta t, \Delta x; y)=\frac{\langle v(x, y, z, t)u(x+\Delta x, y, z, t+\Delta t)
\rangle_{x,z,t}}{\langle v(x, y, z, t)u(x, y, z, t)
\rangle_{x,z,t}}\,.
\label{cuv1}
\end{equation}
The averages are over time and also over all points in an $x$-$z$-plane at 
fixed height $y$ (in the model and the DNS). Time correlations at one point are given by
\begin{equation}
\tilde C_{vu}(\Delta t; y) = C_{vu}(\Delta t, 0; y) =
\frac{\langle v(x, y, z, t)u(x, y, z, t+\Delta t)\rangle_{x,z,t}}
{\langle v(x, y, z, t)u(x, y, z, t)\rangle_{x,z,t}}\,.
\label{cuv2}
\end{equation}
In order to quantify the asymmetry effects we introduce the following  
measure for the temporal cross-correlations
\begin{equation}
Q_{vu}(\Delta t)=\frac{\tilde C_{vu}(\Delta t)-\tilde C_{vu}(-\Delta t)}
{\tilde C_{vu}(\Delta t)+\tilde C_{vu}(-\Delta t)}\,,
\label{asymmetryratio}
\end{equation}
(the dependence on height is suppressed in these expressions). 
For the extended correlations due to the non-normal amplification
we expect
$|\tilde C_{vu}(\Delta t)| > |\tilde C_{vu}(-\Delta t)|$ such that
$Q_{vu}>0$ for these cases.

The outline is as follows. In the next section, the low-dimensional 
model of a turbulent 
shear flow by Moehlis {\it et al.} (2004, 2005) is discussed. 
Section 3 describes the analysis of the DNS in a nearly 
homogeneous shear flow followed in section 4 by the discussion 
of boundary layer experiments of Knobloch \& Fernholz
(2004). Finally, a summary and an outlook are given.      

\section{Low-dimensional model of turbulent shear flow}
\begin{figure}
\begin{center}
\epsfig{file=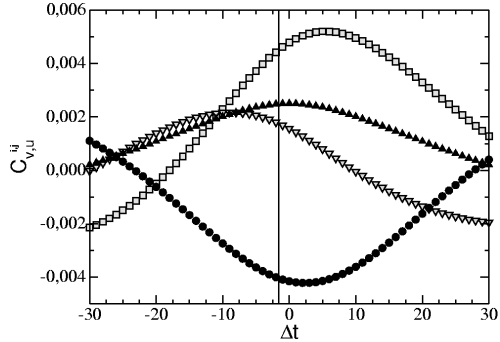,width=6.5cm}
\end{center}
\caption[]{The time correlation function for the four modal
pairs that contribute to the cross-correlation function 
in the nine-mode model: 
${\color{gray}\blacksquare}: C_{vu}^{8,6}$, ${\color{gray}\blacktriangledown}: 
C_{vu}^{8,7}$, $\blacktriangle: C_{vu}^{8,8}$, $\bullet: C_{vu}^{3,2}$}
\label{9modeseparate}
\end{figure}
\begin{figure}
\begin{center}
\epsfig{file=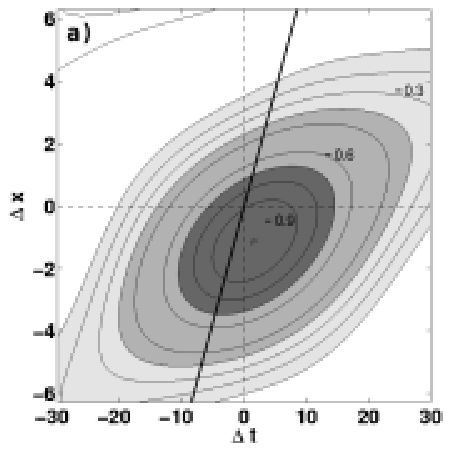,width=6.5cm}
\epsfig{file=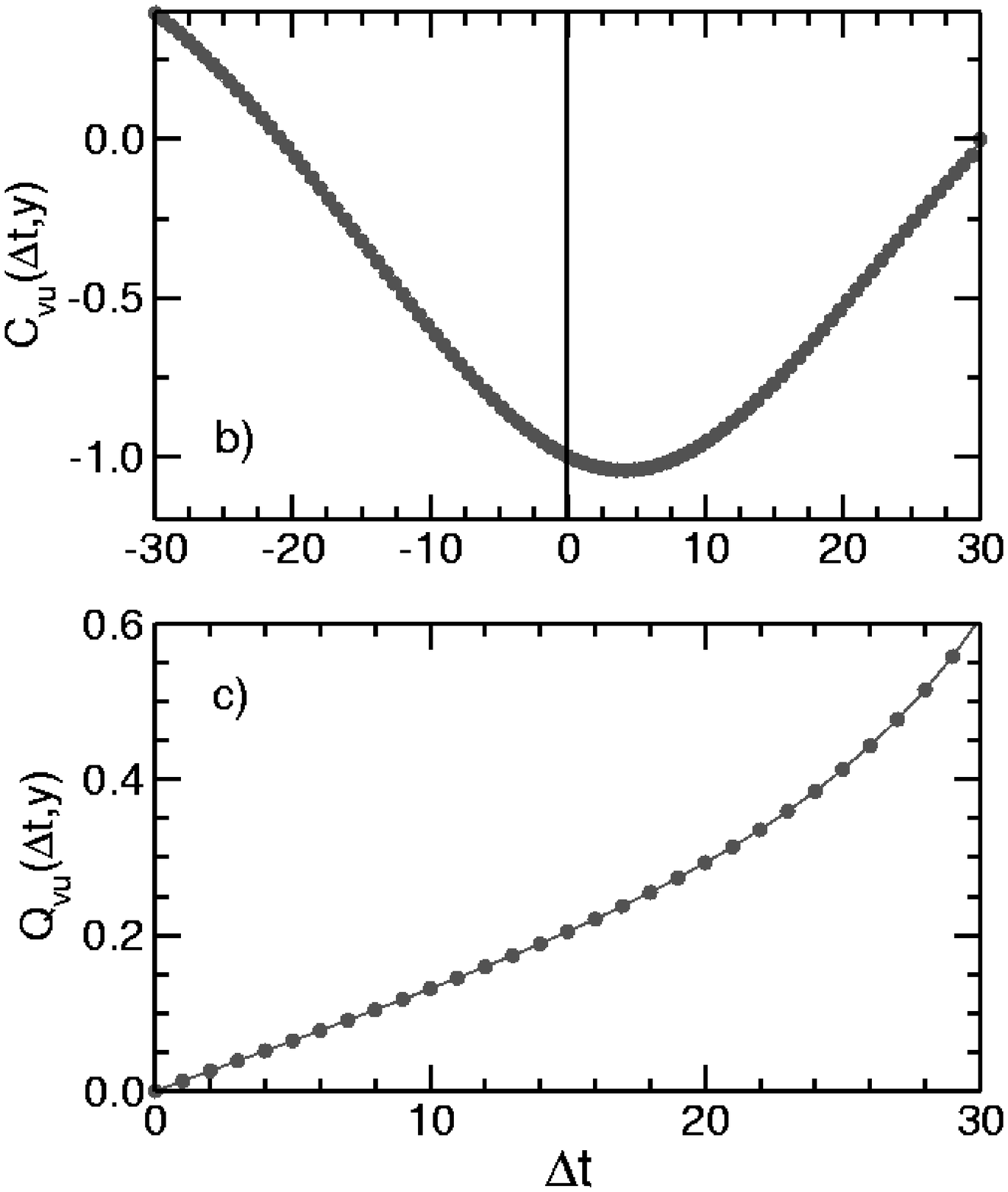,width=5.5cm}
\end{center}
\caption{(a) Space-time cross-correlation $C_{vu}(\Delta t,\Delta x;y)$ in the 
low-dimensional model taken at a height 
$\tilde{y}=y/(d/2)=0.52$. The 
contour levels increase in steps of 0.1. The solid line
represents the dimensionless mean velocity 
$U(\tilde{y}=0.52)=0.74$.
(b) A cut through the contour plot in (a) at $\Delta x=0$ gives the 
temporal cross-correlation $\tilde{C}_{vu}(\Delta t;y)$ .
(c) Asymmetry measure $Q_{vu}(\Delta t)$ corresponding to diagram (b). This measure is 
independent of $y$ due to the spatial mode dependencies.}
\label{vspec}
\end{figure}
The linear model of Eckhardt \& Pandit (2003) can be extended to a
nine-dimensional representation of shear flows, as discussed in more
detail in Moehlis {\it et al.} (2004). The system is confined between two
free-slip planes and driven by a volume force that sustains a laminar sinusoidal
flow profile. The model captures the non-normal amplification process
and completes it with modes for transversal shear and instabilities of the streamwise
streaks. With $L_y=d/2$, the
aspect ratio is $L_x:L_y:L_z=2\pi:1:\pi$, and we simulate the flow at a  
Reynolds number $Re=U_0 d/(2\nu)=180$, where the reference value for the
velocity $U_0$ is determined from the sustained laminar velocity profile at $y=d/4$ 
with $d$ being the distance between the free-slip planes. 

With the Galerkin modes ${\bf u}_i({\bf x})$ and the amplitudes $a_i(t)$
we can write the turbulent velocity field as
\begin{equation}
{\bf u}({\bf x},t)=\sum_{i=1}^N\,
a_i(t){\bf u}_i({\bf x})\,.
\label{galerkin}
\end{equation}
The spatial part of the cross-correlations can be calculated analytically from the
prescribed modes ${\bf u}_i({\bf x})$. The temporal part follows from the 
numerical solution of a system of ordinary differential equations for the $a_i(t)$
that results with (\ref{galerkin}) from the Navier-Stokes equations.
Equation (\ref{cuv1}) then becomes
\begin{equation}
C_{vu}(\Delta t, \Delta x; y)=\sum_{i,j=1}^N \langle a_i(t) a_j(t+\Delta t)\rangle_t\,
\langle v_i(x, y, z)u_j(x+\Delta x, y, z)\rangle_{x,z}\,.
\label{cuv3}
\end{equation}
Of the 45 possible pairs of modes that could contribute to the correlation
function only four terms, involving five modes, do. 
The modes that
contribute are: ${\bf u}_2$, a streamwise 
streak with no streamwise variations; ${\bf u}_3$, a streamwise vortex; 
${\bf u}_6$ and ${\bf u}_7$, which describe two different wall-normal vortices; 
and ${\bf u}_8$, a mode that depends on all three coordinates. 
Thus $C_{vu}=C_{vu}^{3,2}+C_{vu}^{8,6}+C_{vu}^{8,7}+C_{vu}^{8,8}$ where 
the superscripts indicate the particular mode couplings.
Of these correlators, $C_{vu}^{3,2}$ is exactly the one that probes the 
relation between streamwise vortices and streaks and therefore should have 
a significant variation with respect to $\Delta t$. This is indeed the case,
as Fig.~\ref{9modeseparate} shows.

When all contributions are collected, the 
space-time contours for $C_{vu}(\Delta t, \Delta x; y)$   
at $y/(d/2)=0.52$ are obtained (Fig.~\ref{vspec}).  
The time correlations at a fixed position correspond to a cut
along $\Delta x=0$ (cf. Fig.~1b) and the space correlations for fixed time
from a cut at $\Delta t=0$.  The time correlation function is
negative, as is to be expected for a flow where the streamwise velocity
increases in the positive $y$-direction. It is asymmetric and the
asymmetry gives a positive $Q$ in the center of the layer. 
Because of the small number of modes in the system a more detailed dynamical
system study is possible.  
An analysis of the periodic orbits
in Moehlis {\it et al.} (2005) shows that some of them clearly follow
the vortex-streak-instability sequence, but several do not.
Nevertheless, the correlation functions in Fig.~\ref{vspec} show
that the temporal asymmetry expected from the non-normal
amplification process persists even after taking time averages. 

Besides this similarity, there are differences to the linear model.
The contour levels (Fig.~1a) show a global minimum that is shifted from the origin 
toward $(\Delta t,\Delta x)\approx (1,-1)$ for this particular height. 
The shift in time stems from the streak-vortex coupling 
contributions $C_{vu}^{3,2}$ and $C_{vu}^{8,6}$, and is compatible 
with the stochastic model (Eckhardt \& Pandit 2003).
The one in position can be traced back to the coupling of mode 8 with itself,
$\langle v_8(x, y, z)u_8(x+\Delta x, y, z)\rangle_{x,z}=
\pi/(5+\pi^2)\sin(\Delta x/2)\sin(\pi y)$ for $-1<y<1$. 
The space-time contours are elongated along an axis whose slope
has dimensions of velocity. This velocity is not the mean velocity 
at the height of the measurement. As we will show below, 
this is due to an asymmetry in the width of the structures
in the normal direction.

\section{Nearly homogeneous shear flow}
The direct numerical simulations of a turbulent shear flow also refer
to a flow bounded by two parallel free-slip plates, driven by a volume
force that sustains a linear shear flow in the mean, 
$U(y)=Sy$, except for a small boundary layer near the plates. Details of 
the numerical simulations are given in 
Schumacher \& Eckhardt (2000) and Schumacher (2004). 
Relevant parameters for the simulation are listed in table 1.
\begin{table}
\begin{center}
\begin{tabular}{lcccccccc}
     & $N_x\times N_y\times N_z$ & $L_x:L_y:L_z$  & $U$ & $S$    & $\epsilon$ & $u_{rms}$ & $S^*$ & $R_{\lambda}$ \\ \hline
DNS-1 & $128\times 65\times 128$  & $2\pi:1:2\pi$  & $1$            & $2$    & 0.04 & 0.37& 6.1 & 79 \\
DNS-2 & $256\times 129\times 256$ & $2\pi:\pi:2\pi$& $3/\pi$        & $6/\pi$& 0.44 & 1.08& 5.0 & 166 \\
\end{tabular}
\label{tab1}
\caption[]{Parameters of the two DNS that were taken for the analysis. 
$U$ is the mean streamwise
velocity at the free-slip boundary. The mean energy dissipation rate is given by 
$\epsilon=15 \nu \langle(\partial u/\partial x)^2\rangle$, as in
the experimental determinations.
$S=dU/dy$ is the constant  shear rate. 
The root mean square velocity is given by 
$u_{rms}=\langle u^2\rangle^{1/2}$, the dimensionless 
shear parameter is $S^*=S u_{rms}^2/\epsilon$ and the 
Taylor microscale Reynolds number 
$R_{\lambda}=\sqrt{15/(\nu \epsilon)}\,u_{rms}^2$. The spectral resolution 
criterion $k_{max}\eta >1$ is satisfied with $k_{max}=\sqrt{2}N_x/3$.} 
\end{center}
\end{table}

\begin{figure}
\begin{center}
\hfill
\epsfig{file=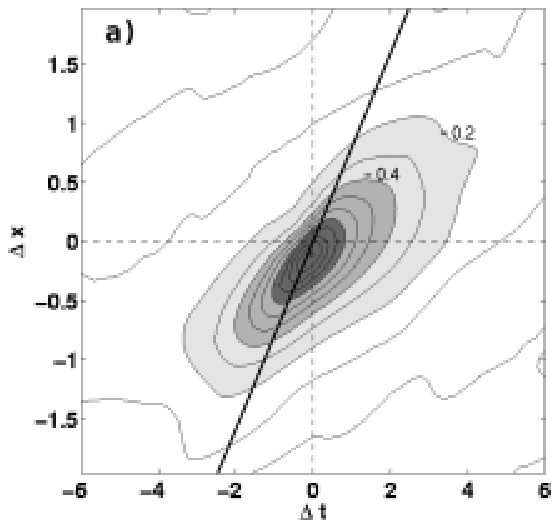,width=6.5cm}
\hfill
\epsfig{file=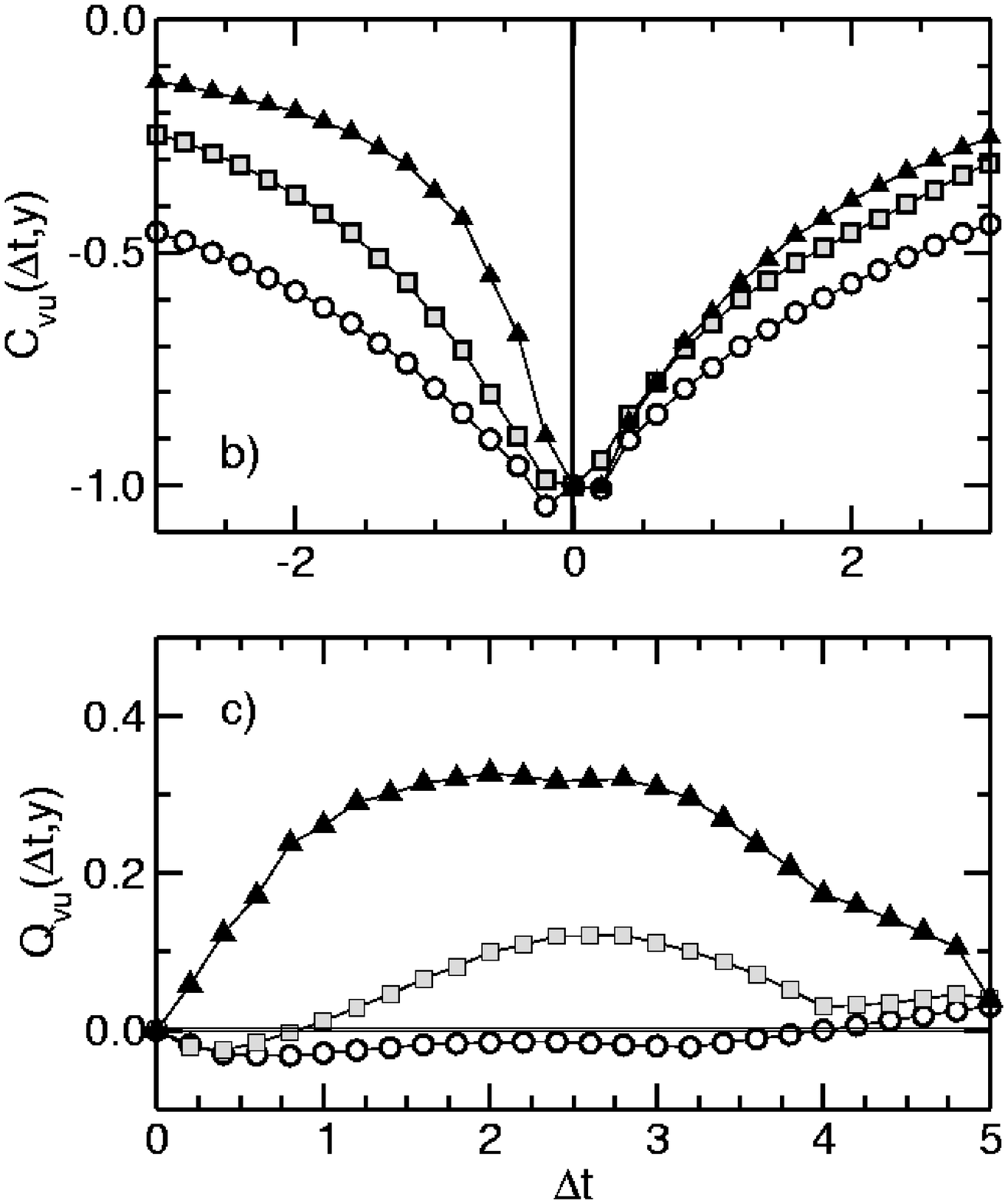,width=5.5cm}
\hfill
\end{center}
\caption[]{Space-time cross-correlation $C_{vu}(\Delta t, \Delta x; y)$ of a nearly homogeneous 
shear flow. Data are  taken from DNS-1 (see table~1) at a height 
$y/L_y=0.11$. The Reynolds number $Re=U L_y/\nu=1800$ where $U$ is the 
mean turbulent velocity at the boundary. 
(a) Space-time plot of the cross-correlations. The 
contours increase in steps of 0.1 and the unit value is at the origin. The solid line
represents the dimensionless mean turbulent velocity $U(y/L_y=0.11)=0.77$.\\
(b) Temporal cross-correlation along $\Delta x=0$ for different heights: 
$\blacktriangle: y/L_y=0.13$, ${\color{gray}\blacksquare}: y/L_y=0.25$, $\circ: y/L_y=0.50$.
(c) Asymmetry coefficient for $\tilde{C}_{vu}(\Delta t;y)$ at the same heights as (b). }
\label{C_DNS}
\end{figure}

Space-time contours for the cross-correlations in run DNS-1 are shown 
in Fig.~\ref{C_DNS}a. Several features are similar to the ones in 
the low-dimensional model: it has the same asymmetry with 
respect to time and the iso-countours are oval and not aligned with the
coordinate axis. 
Thus, even though more spatial degrees of freedom are present,
the Reynolds number is higher, and the turbulence is fully
developed, the non-normal amplification is reflected in the 
Eulerian cross-correlation function. 

However, there are noticable additional features. The asymmetry
measure shows a pronounced height dependence, being strongest
close to the bounding surfaces and getting weaker towards the
center. In addition, it shows a time interval where its value
is negative, $Q_{uv}<0$ (see Fig.~\ref{C_DNS}c). This interval is almost negligible
close to the walls and becomes longer as the reference position
moves towards the center. 
 We see this phenomenon linked to the difference in the number of
modes that can contribute, and hence to the possibility of 
additional dynamical processes. Assuming that the smallest scale
is set by dissipation and does not vary much across the flow,
the largest scale for the possible structures is set by the distance
to the free-slip boundary. By this reasoning there are fewer active modes
close to the wall than in the center, thus limiting the nonlinear
interactions and highlighting their correlations.

\begin{figure}
\begin{center}
\epsfig{file=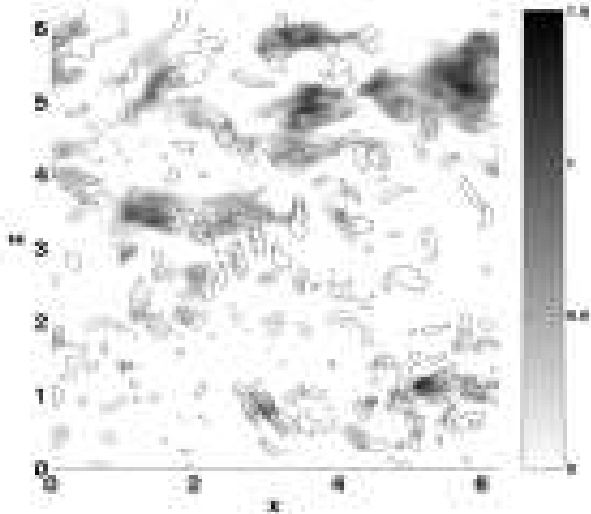,width=8cm,height=7cm} 
\end{center}
\caption[]{Snapshot of the two turbulent velocity fields 
entering the cross-correlation function.
A slice cut from a sample of DNS-2 (see table~1) is taken. The turbulent streamwise 
velocity $u$ is indicated by shading for values of $u$ between $0$ and $1.5$, only. 
The contour lines show three isolevels 
of the wall-normal component $v$ at values of $-0.8$, $-0.5$ and $-0.2$. 
The maxima are shifted relative to each other by a small streamwise distance
that corresponds with the shift of the maximum of the space-time cross-correlation by 
$\Delta x$ as visible in Fig.~3a.}
\label{uv}
\end{figure}

A spatial plot of the streamwise turbulent velocity 
component $u$ and the shear component $v$ in the
$x$-$z$ plane for one point in time and at a fixed height $y$ in the
layer (see Fig.~\ref{uv}) reveals that the
contributions to the cross-correlation function come from
fragmented regions, of an extension compatible with the dimensions
of coherent structures. Negative contours of $v$ 
indicate streamwise vortices which generate the streamwise 
streaks (shown as gray-filled contours of $u$). The maxima
of $u$ and $v$ contours are displaced slightly, 
in accordance with the off-set in the maximum of the spatial 
cross-correlation in Figs.~2a and 3a. 

The off-set can be explained by the observation that 
a streamwise vortex pair centered at height $y$ will 
be advected with the corresponding
mean streamwise velocity at that height, $U(y)$.
The pair will mix slower 
moving fluid into a region that moves on average faster,
and hence will temporarily reduce the local advection velocity $U_c$ 
to values below the mean velocity
$U(y)$. But since $U_c$ will advect the streamwise streak that is about 
to be lifted up, it will remain behind 
the vortex pair, resulting in the spatial shift of the 
most intense cross-correlation. 

\begin{figure}{bt}
\begin{center}
\epsfig{file=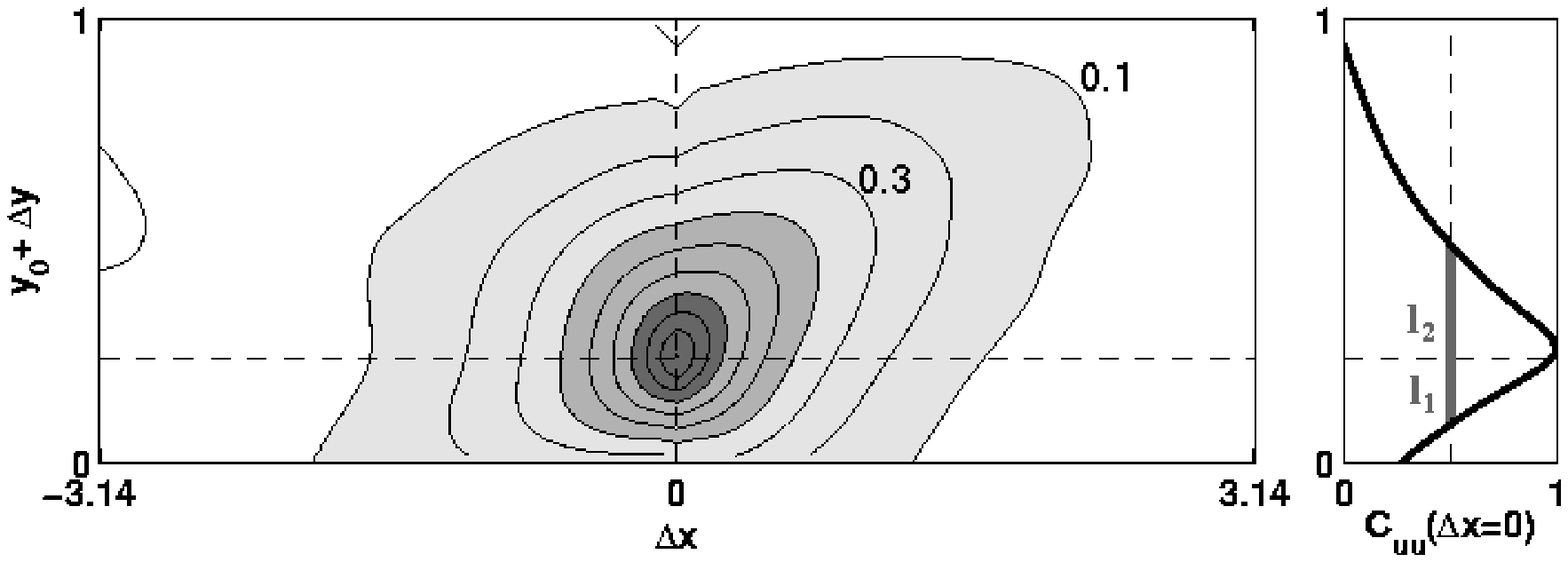,width=7.5cm,height=3cm}
\epsfig{file=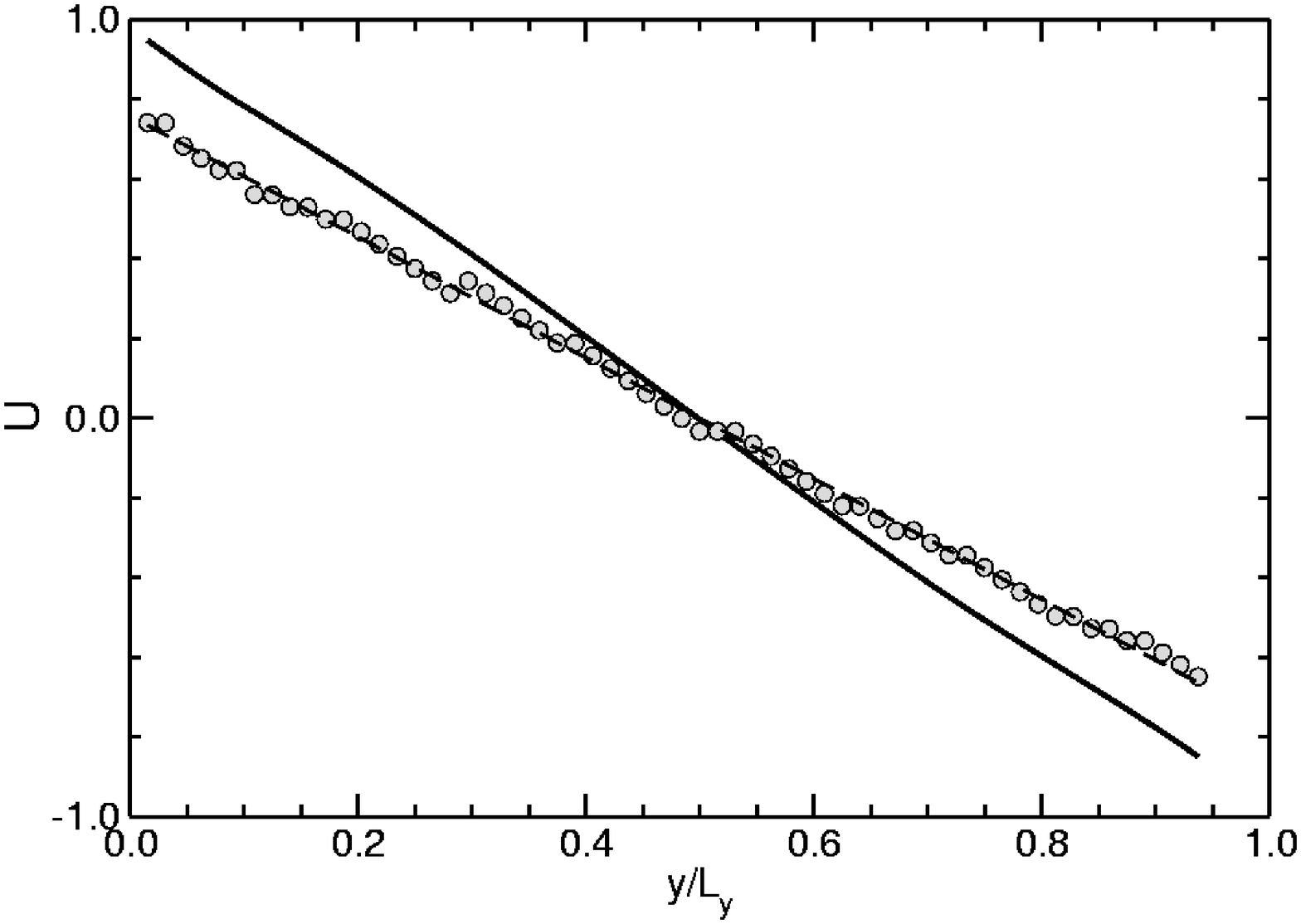,width=5.5cm}
\end{center}
\caption[]{Left: Contour plot of 
spatial autocorrelation function $C_{uu}(\Delta x,\Delta y; y_0)$ as given by (\ref{cuu})
in streamwise and wall-normal
directions taken at $y_0=1/4$ from DNS-1. 
The inclination of this structure is about 9 degrees.
Mid: A cut through the contour at $\Delta x=0$ is shown. 
It is also indicated
how the two widths at half of the maximum, $\ell_1$ and $\ell_2$, 
of the resulting autocorrelation function are defined. Right: 
Comparison of the mean $(U)$ and the convection $(U_c)$ 
velocities for data from DNS-1.
Black solid line is for mean velocity as a function of the 
wall-normal distance $y$. 
The symbols stand for the corresponding convection 
velocity as given by (\ref{Uc}). } 
\label{uu}
\end{figure}

The correlation functions shown in Fig.~2 and 3 
and many others for different aspect ratios and Reynolds number 
show an inclination of the
isocountours in the spatio-temporal correlations
$C_{vu}(\Delta t, \Delta x; y)$. Since the two axes being compared have
dimensions of time and length, the inclination has the dimension of a
velocity: but as the comparison with the straight lines in both
figures shows, the velocity with which these structures are advected
is systematically lower than the mean velocity at that height, $U(y)$. 
We can trace this effect back to an asymmetry of the spatial autocorrelation 
function of the streamwise turbulent velocity in the normal direction, as measured by
\begin{equation}
C_{uu}(
\Delta y;y_0)=
\langle u(x,y_0,z,t) u(x
,y_0+\Delta y,z,t)\rangle_{x,z,t}\,.
\label{cuu}
\end{equation}
The left and middle diagram of Fig.~\ref{uu} show that this 
function is asymmetric with respect to the wall-normal direction,
obviously influenced by the presence of the free-slip walls at $y=0$ and $y=L$. 
If we estimate the mean advection speed of the coherent structures from
an average of the streamwise speed over a domain determined by the 
full width at half maximum of $C_{uu}(\Delta x,\Delta y;y_0)$, we find
\begin{equation}
U_c=\frac{1}{\ell_2-\ell_1}\int^{\ell_2}_{\ell_1} 
U(y)\,\mbox{d}y\,. \label{Uc}
\end{equation}
Here, $\ell_2$ and $\ell_1$ are the widths at half maximum of the asymmetric 
$C_{uu}(\Delta x=0,\Delta y;y_0)$ (see also the mid diagram of Fig.~\ref{uu}). 
The result of such an averaging procedure can be seen in the right diagram of 
Fig.~\ref{uu}. The convection velocity as defined by (\ref{Uc}) becomes smaller as the 
mean velocity and coincides
with the inclination of the space-time contours of the velocity cross-correlations 
of Fig.~3. The coherent structures thus move with the streamwise speed as
determined by an average over their size.

\section{Turbulent boundary layer}
The third class of flows for which we determine cross-correlation functions are high-Reynolds number boundary layer flows. 
The experiments were done at a wind tunnel of the Hermann-F\"ottinger
Institute (HFI) in Berlin and the German-Dutch Windtunnel (DNW).
Triple hot-wire probes allowed measurements of all three velocity 
components. With sampling rates of 20 kHz at HFI and 125 kHz at DNW,
data sets containing about a million data points at the HFI and 5 million at the DNW
could be obtained. The boundary layer thickness (see the
footnote on p.~3) was found to be 
$\delta=63$mm at HFI and $\delta=240$mm 
at DNW.  Some parameters are summarized in table 2; further 
experimental details may be found in 
Knobloch \& Fernholz (2004).

\begin{figure}
\begin{center}
\epsfig{file=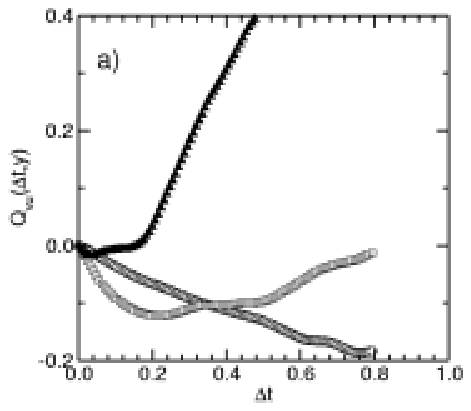,width=6cm}
\hspace{0.5cm}
\epsfig{file=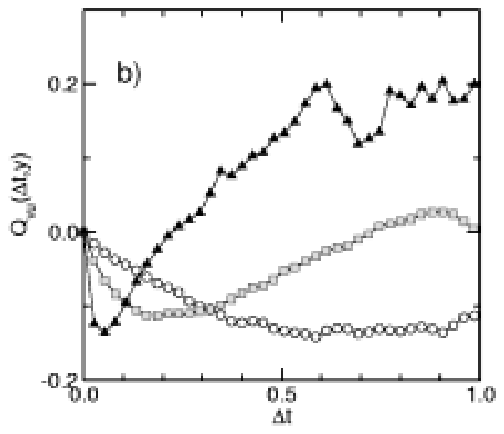,width=6cm}
\end{center}
\caption[]{Asymmetry coefficient 
$Q_{vu}(\Delta t)$ for two sets of turbulent boundary layer data.  Time is given in units
of $\delta/U_{\infty}$ for both figures. (a): HFI measurement at
$Re_{\delta}=41600$ for three different heights $\tilde{y}=y/\delta$ above the wall,
$\blacktriangle: \tilde{y}=0.02$, ${\color{gray}\blacksquare}: \tilde{y}=0.11$ and $\circ: \tilde{y}=0.31$.
(b): DNW measurement at Reynolds number $Re_{\delta}=1237900$.
Here $\blacktriangle: \tilde{y}=0.02$, ${\color{gray}\blacksquare}: \tilde{y}=0.11$ and $\circ: \tilde{y}=0.34$
(see also table 2 for more details).}
\label{experiment}
\end{figure}

Because of the measurements at a single location, only
the short time behaviour of the cross-correlation functions can be
determined. The results for different distances from the wall and 
different Reynolds numbers are shown in Fig.~\ref{experiment}.
Already for the data set closest to the wall there is a time interval
with negative values in the asymmetry measure, and this
interval increases as one moves further out. 
Comparison
with the homogeneous shear flow DNS suggests that this point
is already in the transition region away from the 
vortex dominated near wall layer.
For the points furthest 
from the wall no reversal to positive values is detected.

The data from HFI and DNW are collected at about the same 
relative positions when heights are measured in units of 
the boundary layer thickness, $y/\delta$. 
The asymmetries at these heights
show remarkably similar behaviour,
especially for the value $\tilde y= y/\delta=0.11$, which 
is present in both data sets, and for the two similar
values $\tilde y=y/\delta=0.31$ (HFI) and $0.34$ (DNW).
In wall units, the $\tilde y=0.11$ 
corresponds to a height of 
$y^+\sim 174$ for the HFI data and $y^+\sim 3950$ for the DNW data. 
In agreement with the findings of \cite{dGE00} and 
\cite{DelAlamo03} for turbulent intensities, 
the cross-correlations collapse
in external scaling, i.e. relative to boundary layer thickness 
and the external velocity. 
A possible explanation for this behaviour could be
that the intermittent bursting activity in the boundary layer
lifts fragments of the coherent structures 
higher into the intermediate layer, where their further breakup is 
determined by the boundary layer thickness $\delta$ 
(see e.g. Blackwelder \& Kovaszany 1972). Analysis of additional
data shows that the asymmetry in the time correlations shows
up for positions up to $y=0.05 \delta$.

\begin{figure}
\begin{center}
\epsfig{file=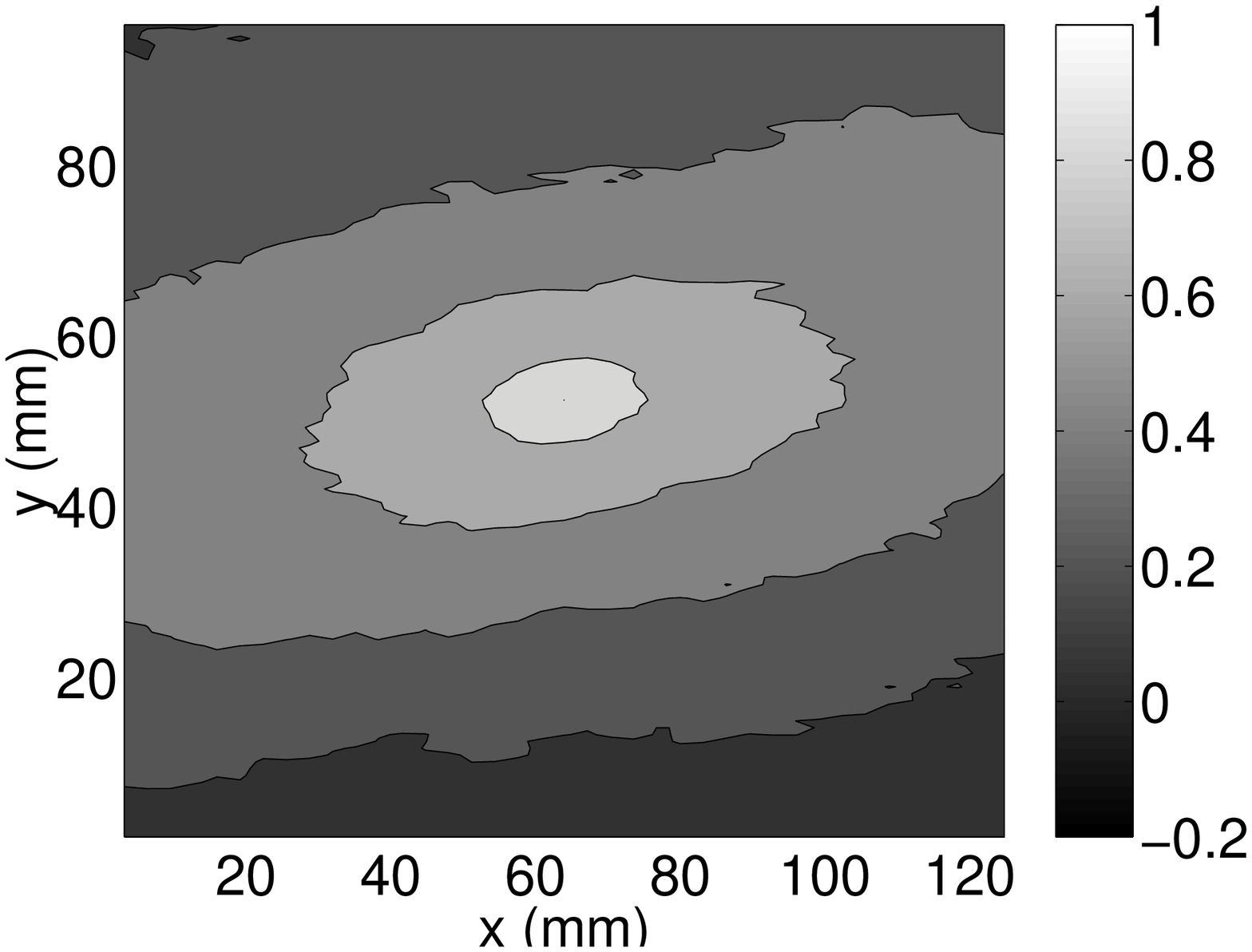,width=0.6\textwidth}
\end{center}
\caption[]{Contour plot of 
spatial autocorrelation function $C_{uu}(\Delta x,\Delta y; y_0)$ as 
given by (\ref{cuu})in streamwise and wall-normal
directions taken at $y_0/\delta=0.21$ from PIV measurements at the DNW. 
The inclination of this structure is about 11 degrees and similar to 
that in the boundary layer.}
\label{cuu_exp}
\end{figure}

Information about the instantaneous in-plane correlations similar to
Fig.~\ref{uu}a can be obtained from PIV measurements at fixed heights. The 
data in Fig.~\ref{cuu_exp} show that the inclination in 
$C_{uu}(\Delta x,\Delta y; y_0)$ is preserved and has about the same value.

\begin{table}
\begin{center}
\begin{tabular}{lcccccccc}
     & $y/\delta$ & $y^+$ & $U_{\infty}$ & $S$ & $\epsilon$ & $u_{rms}$ & $S^*$ & $R_{\lambda}$ \\ 
     &            &       & $[m/s]$      & $[1/s]$ & $[m^2/s^3]$& $[m/s]$ &     & \\ \hline  
HFI  & 0.02 & 34    & 10 & 1576 & 39.6   & 0.97 & 37.8 & 151 \\
     & 0.11 & 174   & 10 & 143  & 10.6   & 0.75 & 7.5  & 172 \\
     & 0.31 & 519   & 10 & 72   & 4.7    & 0.64 & 6.4  & 192 \\
DNW  & 0.02 & 709   & 80 & 1447 & 1146.4 & 6.31 & 50.3 & 1156 \\
     & 0.11 & 3953  & 80 & 242  & 320.6  & 5.35 & 21.6 & 1574 \\
     & 0.34 & 12507 & 80 & 109  & 123.6  & 4.27 & 16.1 & 1614 \\
\end{tabular}
\label{tab2}
\caption[]{List of boundary layer measurements. 
We have picked three distances from
the wall for every free-stream velocity $U_{\infty}$. $y^+=y u_{\tau}/\nu$ with 
$u_{\tau}=(\tau_{wall}/\rho)^{1/2}$. 
The other quantities are defined as in the caption of Table~1.}
\end{center}
\end{table}


\section{Summary}
A comparison between the three sets of data shows that the vortex-streak
interaction is reflected most strongly in the cross-correlation function 
closest to the wall or to bounding surfaces. Further away the signal gets
weaker, as expected by the change in structures (Robinson 1991).
The similarity of the asymmetry measure further out suggests the
prevalence of a similar dynamics. Interestingly, the asymmetry measures 
are similar when the height is measured in units of boundary layer thickness,
rather than viscous length scales. It will be interesting to see
how horsehoe vortices and their dynamics are reflected in correlation
functions, and whether they can explain the cross-correlation functions
or whether other dynamical processes have to be identified.

\begin{acknowledgments}
The work was supported by the Deutsche Forschungsgemeinschaft (DFG)
within the Interdisciplinary Turbulence Initiative, and by the DAAD within 
the PROCOPE program. We thank the John von Neumann Institute for Computing
at the Forschungszentrum J\"ulich for continued computer access and support.
\end{acknowledgments}

\end{document}